\documentclass[a4paper,10pt]{article}

\usepackage[pdftex]{color,graphicx,hyperref}
\usepackage{amsmath,amsfonts,amssymb}
\usepackage{algorithm}
\usepackage{algpseudocode}
\usepackage[margin=1in]{geometry}
\usepackage{setspace}
\usepackage{booktabs}
\usepackage[labelfont=bf,labelsep=period]{caption}
\usepackage{subcaption}
\usepackage{scicite}
\usepackage[utf8]{inputenc}
\usepackage[affil-it]{authblk}
\usepackage{comment}
\usepackage{times}

\title{Multiplexity is temporal: effects of social times on network structure}

\author[1*]{Javier Ureña-Carrion}
\author{Sara Heydari}
\author{Talayeh Aledavood}
\author{Jari Saramäki}
\author{Mikko Kivelä}

\affil{\small{Department of Computer Science, Aalto University School of Science, 00076 Aalto, Finland}}
\affil[*]{\small{Corresponding author email: javier.urenacarrion@aalto.fi}}

\date{}

\begin{document}

\maketitle

\begin{abstract} 
Large-scale social networks constructed using contact metadata—such as emails, texts, and calls—have been invaluable tools for understanding and testing social theories of society-wide social 
structures. However, multiplex relationships explaining different social contexts have been out of reach of this methodology, limiting our ability to understand this crucial aspect of social systems.
We propose a method that infers latent social times from the weekly activity of large-scale contact metadata, and reconstruct multilayer networks where layers correspond to social times. We then analyze the \textit{temporal multiplexity} of ties in a society-wide communication network of millions of individuals. This allows us to test the propositions of Feld's social focus theory across a society-wide network: We show that ties favour their own social times regardless of contact intensity, suggesting they reflect underlying social foci. We present a result on strength of monoplex ties, which indicates that monoplex ties are bridging and even more important for global network connectivity than the weak, low-contact ties. Finally, we show that social times are transitive, so that when egos use a social time for a small subset of alters, the alters use the social time among themselves as well. Our framework opens up a way to analyse large-scale communication as multiplex networks and uncovers society-level patterns of multiplex connectivity.

\end{abstract}

\section{Introduction}

Tie multiplexity is a central notion of social network analysis: the idea that people interact in social contexts that encompass different types of relationships --overlapping combinations of friends, family, work colleagues, and a myriad of acquaintances \cite{Verbrugge1979,MacRae1951}. 
Analyzing real-world social contexts and tie multiplexity can be challenging, as it may require either prohibitively expensive surveys or additional data sources \cite{Saramaki2015,Pescosolido2021}.
Simultaneously, massive auto-recorded contact datasets, such as phone call logs, have been crucial for the development of computational social science and our current understanding of the dynamics of social systems \cite{Saramaki2015,Onnela2007,Miritello2013}. These studies have shed light on diverse patterns of human communication including burstiness and its effect on information spreading \cite{Barabasi2005,Karsai2011,Karsai2012,Miritello2011}, how such patterns can be revealing of tie stability \cite{Navarro2017,VergaraHidd2023}, and strategies and rates at which people communicate with their alters \cite{Miritello2013, aledavood2016,Iniguez2023}. However, although these data are rich in temporal patterns, many sources do not contain any information on the content of the messages between pairs of individuals which could be used to infer the context of the tie. Due to this, for large-scale data, notions of multiplexity have largely focused on different types of interactions on social media  \cite{dickison2016multilayer,celli2010social,omodei2015characterizing,chen2021polarization}. 

Despite the seeming lack of information about multiplexity in communication data, their temporal patterns can indicate the context of the tie, as human behavior follows temporal rhythms influenced by both natural and societal factors. The natural components align with the 24-hour cycles of wakefulness and sleep\cite{Panda2002,Aledavood2015b}, while societal elements manifest in designated times for labour or leisure. Sociotemporal patterns are shaped by cultural and legal contexts, exerting profound effects on behavior and overall well-being \cite{Helliwell2015,Jeong2020}. Earlier studies that combine surveys with contact metadata have identified a preference among individuals for specific contact times depending on the nature of their relationships. For instance, female students tend to contact friends predominantly in the evenings, while their communication with family members encompasses shorter calls at various times \cite{Aledavood2015}. Elderly patients, however, contact health professionals at almost any time but tend to favour daytime for family and evenings for friends \cite{Aubourg2020}. On the other hand, weekend activity is largely absent in datasets of work emails \cite{Aledavood2015b}. 
Although individuals may contact one another at any point throughout the week, we could expect contact times to generally align with broader societal rhythms, depending on the nature of their relationship. Previous research has shown that activity during weekends is associated with higher tie strength \cite{Urena2020}. More recently, Vergara Hidd \textit{et al.} showed a connection between the stability of relationships and weekend activity, highlighting that transient relationships (that begin and end within a few months) tend to initiate and remain active predominantly during weekdays. In contrast, longer-term connections communicate during the weekends as well, which can serve as a proxy to know whether a new relationship will become stable \cite{VergaraHidd2023b}.  

\begin{figure}
    \centering
    \includegraphics{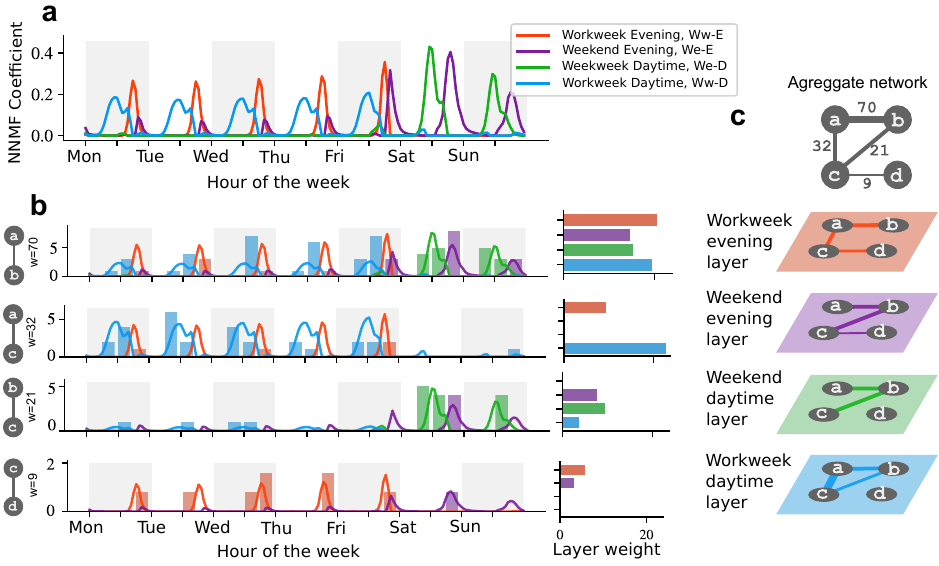}
    \caption{\textbf{Social times as a basis for multiplexity.} Different ties are active at different times. Using NNMF we obtain population-level latent signals of social activity and reconstruct a multilayer network based on how they explain the calling patterns of each tie. In our approach, all nodes (people) are in all layers, but ties (relationships) can be present in one, some or all layers. \textbf{(a)} We obtain social times by decomposing the weekly activity matrix of the population using NNMF. The plot depicts the signal strength for each hour of the week. We name signals based on the times when they are strongest. \textbf{(b)} Each tie can be modeled as a linear combination of different signals. For four example ties (rows), we show a histogram of the number of contacts placed during the week (each bin is 6 hours, with $w$ total contacts). Each tie is described by a unique linear combination of social times, or layer weights. The stylized densities (lines, scaled to match histogram counts) depict tie-specific weights for each layer. \textbf{(c)} Visual depiction of how an aggregate network of four ties can be reconstructed as a multilayer network, with ties having different weights, if any, for each social time. As an example, tie \textit{a-b} is present in all layers as all coefficients are positive, while tie \textit{a-c} is only active in work-week layers.}
    \label{fig:reconstruction}
\end{figure}

We propose a method for inferring the \textit{temporal multiplexity} of a tie from communication data, where we leverage population-level behaviour as a basis for social contexts. 
We then apply this methodology to a dataset of Call Detail Records (CDRs) from a large European country to construct and analyse a society-wide multiplex social network.
Our core assumption is that categories of ties are associated with particular timings during the week (e.g., work ties will be largely active during work hours), so that information on tie multiplexity is encoded in population-wide activity signals. Such signals are not unique, i.e., there is not a single set of hours that captures working times for the whole population, but they do represent latent components that capture much variation within calling patterns. We use orthogonal non-negative matrix factorization (NNMF)\cite{Kimura2015}, a method commonly used for inferring temporal patterns \cite{cazabet2018,Aledavood2022}, for inferring the population-wide signals. 
We then reconstruct multilayer networks by proposing a statistical model where each layer represents a population-wide signal, i.e. a temporal latent component obtained via the NNMF, and each tie may be present in one, some or all layers. 
We visualize this process in Figure \ref{fig:reconstruction}: we first decompose population-wide activity during the week using NNMF, and then express each tie as the combination of latent societal activity rhythms. This way, each relationship is not only described by its presence in different layers but by how it allocates weights across layers. This method does not classify ties according to clear-cut relationships. Instead, it places relationships in a continuum of temporal components that are informative of social contexts. 

We test our work by following the sociological framework of Feld's focus theory \cite{Feld1981}. Feld proposed that relationships, along with their multiplexity, emerge from social foci —environments like workplaces or otherwise social spaces that foster interactions. Just as auto-recorded communication produces digital traces, the focused organization of ties leads to patterns in the structure of social networks, including clustering and bridging phenomena. While focus theory lacks an explicit temporal component, it proves to be a flexible framework with well-defined propositions regarding network structure. In our analysis, we do not infer particular social foci, but evaluate whether the use of social times corresponds to the expected patterns of focus theory. 

Our results show that social times during the week capture essential aspects of human communication and behaviour. The way that ties balance their contacts across these social times, which we call temporal multiplexity, captures tie-specific behaviour that is largely independent of the amount of contacts and resilient to bursty dynamics. Temporal multiplexity is related to tie strength \cite{Granovetter1973,Onnela2007,Urena2020} as it captures a spectrum of topological bridginess: monoplex ties serve as local bridges between groups while multiplex ties are embedded in overlapping circles of friends. Our results show that the usage of social times is transitive in the sense that triangles are preserved in the layers on conditions dependent on the egos. Last, we find that layers help untangle clustering patterns, as weekend layers tend to contain alters that have more connections within ego networks.

\subsection*{Theory of social foci}

Feld's work is part of a trend in sociology that formulated social theories in terms of networks, which became increasingly prominent in the latter half of the previous century \cite{Pescosolido2021}. In this vein, Granovetter proposed that the micro-level strength of interpersonal ties is crucial for overall macro-level network connectivity: \textit{weak ties} are "strong" in the sense that they act as bridges, and connect people that are otherwise distant \cite{Granovetter1973}. 
Focus theory emphasizes underlying societal structures that lead to social connections. People socialize through repeated interaction within institutional, legal or otherwise social spaces, so their connections are not random \cite{Feld1981}. A social focus impacts ties depending on its size and its constraining capacity, i.e., the number of people it includes and the amount of interaction it promotes. For example, a university is both large and not constraining, as any two people are not likely to interact much; a particular class is smaller and more or less constraining depending on how much it promotes discussion or teamwork. Overall, small and constraining foci promote many connections within a group, whereas larger and non-constraining foci favour less dense connections. 

Based on these notions, Feld made several propositions of how foci impact the bridginess of ties, the density of ego networks, among others. A tie is more likely to be bridging if it is associated with few, large and unconstraining foci. Bridging ties can be, e.g., acquaintances that exist within specific social roles, such as a work colleague from a different department. Conversely, a non-bridging tie would be associated with several compatible foci; that is, a relationship that exists within several social spaces. This provides an underlying context for Granovetter's weak tie hypothesis: the bridginess of "weak" ties is related to both their tie strength and monoplexity, they have few overlapping friends as they fulfill a single social role. 

\section{Results}

We use a large dataset of CDRs during the first seven months of 2007 in a European country. CDRs consist of metadata from phone calls including anonymized callers, callees and a timestamp. Our dataset has $\sim 6.5$ million users and $N\sim 10 $ million undirected ties. For each tie, we construct a $24\times7=168$-vector of weekly activity, where each entry contains the number of calls placed during that hour of the week. We represent the population activity on a weekly activity matrix $X$ of size $N\times 168$.

\subsection{Obtaining social times and reconstructing a multilayer network}

The cumulative activity of ties helps us uncover the latent social times. While any given tie is described through their contacts during the week, NNMF unveils social components by describing time in terms of latent signals and describing ties in terms of how they use these signals (see Figures \ref{fig:reconstruction} \textbf{a} and \textbf{b}, respectively). 
We obtain four social times by decomposing a sample of $N_s = 500K$ ties from the matrix of weekly activity profiles $X$ via NNMF with orthogonality constraints \cite{Kimura2015}. More formally, this method approximates large and sparse non-negative matrices as the product of low-rank matrices $H$ and $G$ that can be interpreted in terms of the original data and $J=4$ latent components. The matrix of social times $H$ (of size $168\times J$) captures the strength of the latent components during each hour of the week, while the matrix $G$ of size $N_s \times J$ captures how ties divide their contacts across social times. Most hours of the week have overlapping signals, so orthogonality eases interpretability by favoring independence between the columns of $H$, i.e., by finding signals that are as distinct as possible. We name each of the signals according to the social times when they are stronger: a combination of weekdays and weekends during the daytime and evenings. We selected $J=4$ as previous work has shown it to be optimal for phone usage data \cite{Aledavood2015}, but include latent components for other $J$ values in the Supplementary Material (SM). 

\subsubsection*{Inferring how ties use social times}

We propose an inferential framework to recover the ways ties use social times. This way, latent social times (the columns of $H$) determine contact probabilities, and a tie's contacts are a explained by a tie-specific combination of such probabilities. While it is possible to use the original NNMF formulation to recover tie coefficients of the full dataset, a statistical formulation allows us to succinctly find the layer combinations that best explain the data without over-fitting or having coefficients that are arbitrarily close to zero. In our model, the weekly activity $X_i$ of tie $i$ behaves as a multinomial random variable, which captures the allocation of $w_i$ contacts over $k=168$ hours of the week. The tie-specific probability vector $h_i$ is a $k$-sized linear combination of the latent social times, so that $h_i = h(\alpha_i) = \hat{H}\alpha_i$. Here, $\alpha_i$ is the tie's layer coefficients and $\hat{H}$ is a scaled version of the $H$ matrix to meet probability constraints (details in Methods). Formally, the weekly activity $X_i$ vector of tie $i$ is
\begin{equation} \label{eq:multinomial}
    X_i | \alpha_i, w_i \sim \text{Multinomial}(h(\alpha_i), w_i).
\end{equation}

We developed an algorithm for performing model selection on the coefficients $\alpha_i$ using the Bayesian Information Criterion (BIC). Given the initial approximation of the NNMF classification, we add and delete layer coefficients based on changes in the likelihood function, while ultimately selecting the model that minimizes the BIC (details on inference in Methods, and model section in SM). The activity patterns of ties may be captured by any combination of layers. Since $\alpha_i$ carries the relative proportion of layer weights and thus sums to 1, we define the total weight of tie $i$ on layer $l$ as the product $\hat{w_i} = w_i\alpha_i$. We reconstruct a multilayer network by allocating the estimated layer weights to the ties in our dataset. 
In our reconstructed network, the distribution of ties across layers is largely balanced. With the exception of the weekend daytime layer, which is sparser, most social times (layers) contain between 65-70\% of all ties each. We find that 19.1\% of ties belong to one layer only, and these monoplex ties are equitably distributed across layers --5-6\% of all ties are monoplex and present in each layer. The weekend daytime layer is sparser, with 54.7\% of total ties and less than half the amount of single-layer ties; 2.2\% of all ties are monoplex on weekend daytimes. 

We measure the \textit{temporal multiplexity} of tie as the entropy of its layer coefficients $S(\alpha)= -\sum_{l=1}^J \alpha^l \log(\alpha^l)$. Entropy captures the amount of uncertainty of a system, which in our case means the distribution of contacts across social ties: a monoplex or single-layer tie has zero entropy as all of its contacts can be explained by that layer, and there is no "uncertainty" associated with that tie. Conversely, if all coefficients are equal there is maximum entropy; we can't characterize the tie in terms of single layers or a few layers.

\subsection{Ties favor their own social times}
\label{sec:egotop}

\begin{figure}
    \centering
    \includegraphics{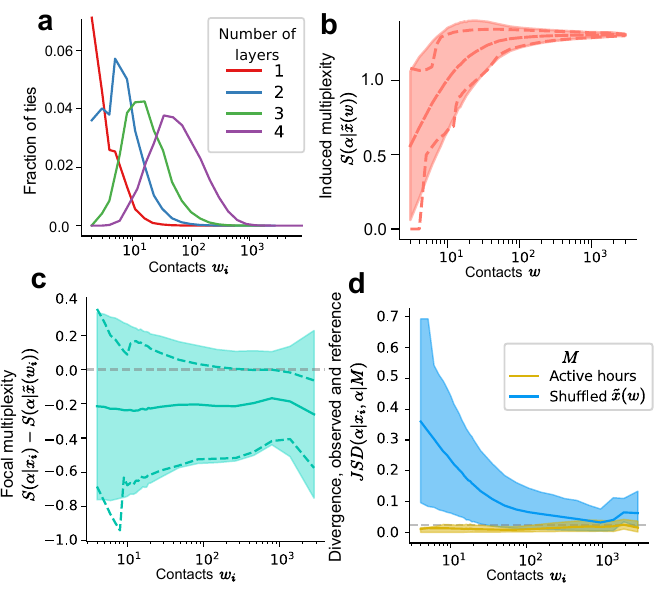}
    \caption{\textbf{Ties favour their own social times.} We test our assumption that social ties have a temporal expression by controlling for other factors that may affect coefficients for social times. 
    \textbf{(a)} The number of layers as a function of the number of contacts shows that low-contact ties tend to be present in few layers, while high-contact ties in many or all layers.  A total of 19.1\% of ties are present in one layer, 29.49\% in two layers, 25.05\% in three, and 26.36\% in all layers.
    \textbf{(b)} We measure the multiplexity induced by contact volume by shuffling calls according to the population activity (fraction of calls during each hour), and classifying ties of $w$ shuffled contacts into layers.
    Higher contacts trivially induce multiplexity, but while low-contacts are heterogeneous, higher-contacts trivially induce a stable level of multiplexity. Mean entropy in solid line, shaded regions capture 1.5 standard deviations, and dotted lines capture the 1st and 9th deciles.
    \textbf{(c)} We define the focal multiplexity as the difference between the observed and the (average) induced multiplexity. The observed multiplexity is consistently lower than the induced (-0.2 units on average) and stable across contact counts. Focal multiplexity displays higher variance than the shuffled profiles, including for high-contact ties.   
    \textbf{(d)} We use JSD to compare the layer coefficients between the full activity profiles and profiles of active hours; and the full activity with shuffled profiles. Shaded regions contain 80\% of the distribution. Using the active hours suffices to broadly capture weight allocation patterns. In contrast, reference weight-allocation patterns under the shuffling model differ substantially from the observed data.
    }
    \label{fig:layeralloc}
\end{figure}

We start by showing that the use of social times is specific to ties and cannot be explained by the volume of contacts or bursty communication. The core idea of temporal multiplexity is that relationships reflect their own social times, which means that they should favor certain social hours more than others. The number of contacts $w$ is strongly associated with multiplexity, such that a large number of calls result in the presence of calls in many social times (see Fig. \ref{fig:layeralloc}a). However, this would also happen by chance if the calls were placed randomly without any preference for one particular layer. What is more, bursty calling patterns \cite{Barabasi2005,Karsai2012} --where consecutive calls are followed by large waiting times--, might result in contacts being concentrated on a layer due to a single, short burst. Our objective is to untangle trivial sources of multiplexity to reveal inherent patterns of time usage within ties. 

We define induced multiplexity as the one that results from shuffling the weekly activity profiles according to the aggregate population behaviour. We denote $\tilde{x}(w)$ as a weekly activity vector of $w$ shuffled contacts (details on Methods). Shuffled activity profiles $\tilde{x}(w)$ have no tie-specific temporal correlations between bins, so that their layer coefficients can't be attributed to latent social timings. We observe patterns of the induced multiplexity $S(\alpha|\tilde{x}(w))$ of shuffled times, which increases along with the number of contacts. While low-contact ties display heterogeneous values, high-contact ties consistently display well-balanced coefficients across layers, with little variance in multiplexity. 

We define the \textit{focal multiplexity} as the difference between the observed multiplexity $S(\alpha|x_i)$, where conditioning on $x_i$ denotes empirical observations, and the induced multiplexity by contact volume $S(\alpha|\tilde{x}(w_i))$. Focal multiplexity is stable across contact counts $w$, showing that induced multiplexity can account for the average increase in observed multiplexity. What is more, focal multiplexity is negative for most ties, i.e., observed multiplexity is consistently smaller than the one induced by contact volume $w$, strongly suggesting that ties favour certain social times more than others. In other words, ties rarely place calls in a way that can be described by the aggregate population-level patterns. Temporal multiplexity is largely tie-specific in the sense it is consistently heterogeneous once we consider the effect of contact counts. 

Next, we assess the effect of burstiness by coarse-graining weekly activity profiles into profiles of \textit{active hours}, binary activity vectors where at least one call was placed at each hour. We compare the classifications of active hours and weekly activity under the premise that the layer coefficients will differ if our methodology is very sensitive to bursty behaviour. 
We use Jensen-Shannon Divergence (JSD) to compare the layer coefficient distributions inferred from the coarse-grained reference ($\hat{\alpha}|AH(x_i)$) and layer coefficients inferred from the observed data ($\hat{\alpha}|x_i$).  JSD is a measure of similarity between probability distributions that takes value zero if the distributions are the same and one if they differ fully \cite{Menendez1997}. Figure~\ref{fig:layeralloc}d show that coefficients are largely aligned (JSD $< 0.03$). This result holds across contact counts $w$, reaffirming that the layer allocation process is robust to differences in calling intensity and burstiness. As a reference, we include the JSD between activity profiles and shuffled contacts. For low-contact ties, the shuffling completely destroys any association between estimated coefficients. These differences decrease for high-contact ties, a likely byproduct of increased entropy in shuffled contact times.

\subsection{The strength of monoplex ties}

The \textit{strength of weak ties}, a central theory in social network analysis, asserts that weak ties are ``strong'' as they act as bridges that provide access to otherwise distant parts of the network, playing a cohesive role in large social systems  \cite{Granovetter1973,Onnela2007,urena2022communication}. In a similar line, Feld proposed that tie bridginess is related to monoplexity \cite{Feld1981}. We assess the effect of temporal multiplexity $S(\alpha)$ on network structure from local and global perspectives. We measure local tie bridginess using topological overlap --a tie-level clustering coefficient \cite{Onnela2007,Urena2020}--, and explore the global emergence of a giant connected component (GCC) when adding ties based on their multiplexity, which we compare with the same phenomenon when adding ties according to their contact volume $w$. In SM, we include additional tests where we quantify changes on shortest path distributions when equal-sized subsets of layers are removed. Taken together, our findings strongly suggest that temporally monoplex ties serve as bridges, with monoplexity capturing an effect that is distinct from the contact counts $w$.

\begin{figure}
    \centering
    \includegraphics[width=.8\textwidth]{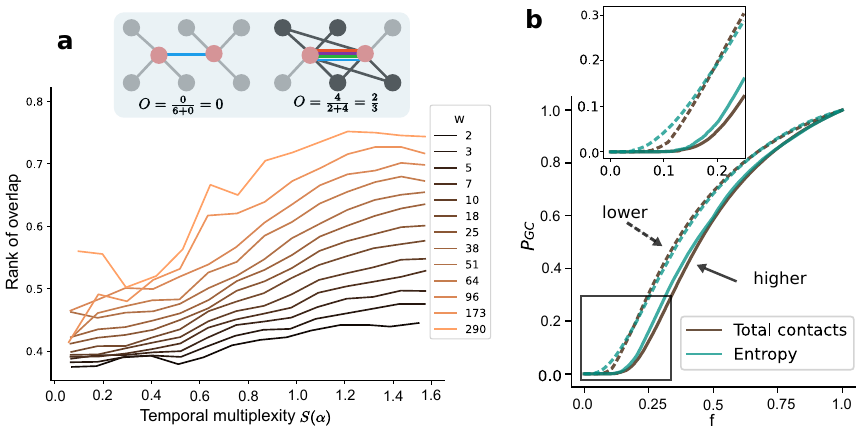}
    \caption{\textbf{Monoplex ties are locally and globally bridging.} We measure multiplexity via entropy, assessing its effect on local tie bridginess and global network connectivity. \textbf{(a)} Within the framework of social foci, monoplex ties (blue link) are more bridging than multiplex ties (multi-colored link). We measure local bridginess with topological overlap $O$, the ratio of common neighbors (dark nodes) to all neighbors (dark and light). 
    (\textit{Bottom}) Ties with lower entropy have consistently low overlap, i.e., they serve as topological bridges, while high-entropy ties have high overlap. These results hold regardless of the number of contacts $w$. Ties are binned based on their contact count, colors depict the lower bin border. \textbf{(b)} Effect of link addition on the appearance of a giant connected component. Starting with all the nodes and no edges, we add links based on total contacts $w$ and entropy, sorted by low-first (dashed lines) and high-first (solid lines). The parameter $f$ denotes the fraction of added links, while $P_{GC}$ denotes the fraction of nodes that belong in the giant connected component. Adding low-entropy ties first leads to the appearance of a GCC faster than adding low-contact ties. After an initial faster appearance, the sizes of GCCs largely align. Both adding high-entropy and high-contact leads to a slower appearance of the GCC.}
    \label{fig:bridginess}
\end{figure}

We measure tie bridginess via topological overlap \cite{Onnela2007,Urena2020}. Given two nodes $a$ and $b$, overlap measures the ratio of common neighbors over all neighbors. More explicitly, for $k_a$ the degree of node $a$ and $n_{ab}$ the number of neighbors common to $a$ and $b$, overlap is $O_{ab}=\frac{n_{ab}}{k_a+k_b-2-n_{ab}}$. 
We measure overlap over the aggregate network as it is indicative of all social connections around a tie regardless of contact times.
Figure \ref{fig:bridginess}a visualizes the core idea of tie bridginess in terms of focus theory: monoplex ties serve as bridges and have low overlap, while multiplex ties exist within communities and have larger overlap. Our results depict the rank of overlap as a function of multiplexity and the number of contacts. For this plot, we calculate entropy on the layer coefficients identified directly via NNMF (see Methods), as enforcing sparsity results in more homogeneous coefficients for low-contact ties. Low-multiplexity ties consistently have lower overlap than high-entropy ties, a result consistent throughout varying levels of contact counts $w$. Even low-contact ties ($w\leq 5)$ are less bridging if they are associated with more layers. In other words, while low-contact ties have been usually considered bridges \cite{Onnela2007}, our results suggest that their temporal expression over the week contains more topological information than the total call count would suggest. Conversely, high-contact ties that display low temporal multiplexity also serve as topological bridges. Our results strongly suggest that temporal multiplexity is aligned with the expectations of focus theory. 

We assess the overall effect of monoplex ties on global network stability through the emergence of the giant connected component (GCC) when adding links. Previous research has shown that low-contact ties behave as weak bridges \cite{Onnela2007}. Since they tend to have few triangles around them, adding weak ties first allows for the creation of paths with few cycles, so low-contact ties have a crucial role in the emergence of a GCC \cite{Onnela2007}. However, our previous results show that even low-call ties can be less bridging locally if they display temporal multiplexity. We compare the effect of adding low-call and low-multiplexity ties first (Fig. \ref{fig:bridginess}b). We find that the GCC appears faster when sorting by multiplexity than sorting by number of contacts, suggesting that monoplexity may serve as a better proxy for overall network connectivity. These differences exist only for the lower values of multiplexity and contact frequency; GCC growth follows a similar pattern for both cases when around $f=20\%$ of links have been added.

\subsection{Social times capture focalized activity around egos}

We explore how social times reflect on individuals by focusing on the topology of ego networks. While our central argument is that the overall distribution of layer weights offers a more comprehensive characterization of social ties, we investigate whether social times hold any structural significance as discrete layers. We do so from two perspectives. First, we see whether egos with larger degrees over- or under-represent their alters in particular layers. We then focus on how the egos divide their alters across social times.
We assess whether alter-allocation is (i) transitive in the sense that the alters are connected in the same layer, and (ii) whether the alters in a social time are embedded in the ego-network, i.e., they have links to other alters regardless of the social time. Inspired by focus theory, we characterize the relative size of a social time as the fraction of alters it contains. At an ego level, a broad social time includes most alters, while a small or confined social time is devoted to a minor fraction of alters. We found this simple statistic to be particularly informative, and we argue that at an ego level such confined layers behave as small and constraining social foci, displaying a high degree of clustering. For completeness, we include additional results on the SM that focus on other topological features.

\begin{figure}[h]   
    \centering
    \includegraphics{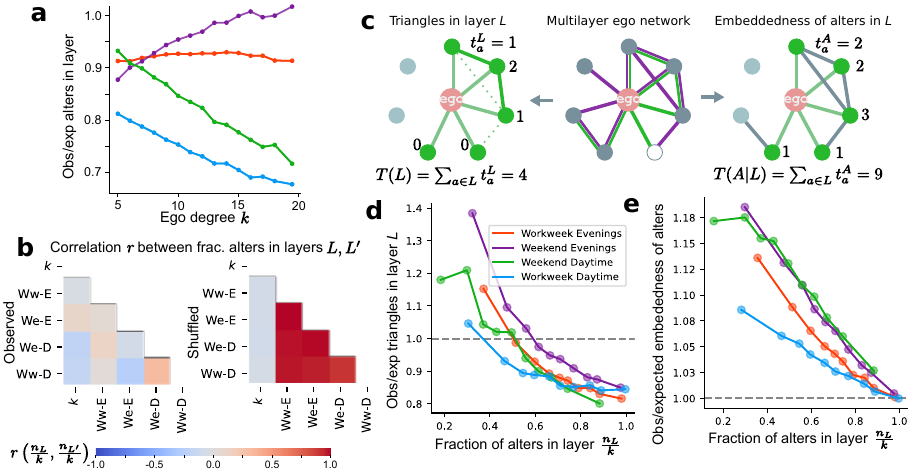}
    \caption{\textbf{Layers capture social behaviour around egos.} We assess whether layers contain relevant topological characterizations around egos.  
    \textbf{(a)} Layers capture some ego-specific behaviour dependent on degrees. We compare the observed and expected number of alters per layer under randomized calls. Higher-degree egos tend to over-represent more alters on weekend evenings, and under-represent alters on daytime layers. \textbf{(b)} Egos allocate alters to layers in unique ways. Pearson correlation for layer sizes $r\left(\frac{n_L}{k}, \frac{n_{L'}}{k}\right)$ show little correlation between the layer sizes. Under the shuffling model, the fractions of alters in layers are highly correlated. \textbf{(c)} We test whether the fraction of alters in a layer is revealing of triangles around an ego. Given the aggregate network $A$ and $n_L$ the number of alters in layer $L$ (in the example $n_L=5$), we sample $n_L$ alters independently (probability dependent on contacts $w_a$). (\textit{Left}) Our first null hypothesis captures the number of triangles $t^L_a$ in a layer $L$ assuming transitivity --if two sampled alters $a$ are connected in the aggregate, they are also connected in the sampled layer; empirically such links might exist in another layer only (dashed line). 
    (\textit{Right}) The triangles around alters $t_a^A$ in the aggregate network $A$ captures whether the alters are embedded in the ego network regardless of the contact time.
    \textbf{(d)} Layers display a negative association between the fraction of alters and the ratio of observed and expected triangles. Egos that contact many of their alters within a layer have less dense personal networks in that layer than expected, but there is also layer-specific behaviour. Weekend layers tend to preserve more triangles, while workweek layers consistently preserve less triangles than expected. Dots denote deciles of $n_L/k$ calculated over the layer.
    \textbf{(e)} All layers have more embedded alters than expected. 
    Weekend layers over-represent such alters at higher rates. Ratios tend to 1 as selecting all alters implies observing all aggregate connections empirically and in expectation.}
    \label{fig:triangles}
\end{figure}

Just as shuffling contact times allowed us to assess the impact of contact counts $w$ on multiplexity, we extend the shuffling model for egos to assess the effect of heterogeneous activity around alters. Here, we shuffle contact times keeping the sequences of alter weights $\{w_a\}_a^k$ fixed. Figure \ref{fig:triangles}a depicts the ratio between the observed and expected number of alters in a layer given the degree $k$. We find layer-specific behaviour, where egos with a larger degree over-represent weekend evening layers. These results are consistent with previous research that has shown that "night owls" --people whose chronotype favour night-times --, tend to have larger personal networks \cite{aledavood2018}.  

We evaluate whether the sizes of social times are correlated for egos, e.g., whether egos consistently communicate with many or few of their alters in two layers. For $n_L$ is the number of alters in layer $L$, and $k$ is the ego's degree, the fraction of alters $n_L / k$ captures the relative size of a layer for an ego. 
As a first step, we compute the correlation between relative layer sizes for both the observed egos and shuffled egos. Figure \ref{fig:triangles}b depicts the correlations between alter fractions in layers and also degree. Our results show slight associations between layers. Particularly, the two daytime layers are positively correlated, and layer sizes are negatively correlated between weekend evening and workweek daytime, but other correlations are minor. A possible explanation for these correlations might lay on the daily rhythms of individuals, who may favor either daytimes or evenings to communicate with their alters \cite{Aledavood2022}. These results, however, stand in sharp contrast with the ego-centric shuffled model, that shows no correlation between layers and degree, and almost complete correlation between layers. The fraction of alters allocated to each social time can't be trivially attributed to the degree or call intensity. 

We evaluate whether social times are transitive, i.e., whether triangles use the same social time. 
For an ego with $n_L$ alters in layer $L$, we evaluate transitivity by comparing the observed triangles in the layer $T(L)$ with the triangles we would observe on a random sample of $n_L$ alters. Our randomized model assumes that transitivity in social times is induced by alters that act independently, i.e., a triangle in the aggregate network is observed in a layer if its two alters are randomly selected, and not because the triad itself favors the social time $L$. We take samples without replacement with probability proportional to the log-weight $\log w_a$ of the alter tie. This sampling strategy addresses possible triangle under-counts: since contact counts tend to be heterogeneous \cite{Heydari2018}, we could expect for a random sample to contain a larger number of low-contact weak ties, which tend to have less triangles around them \cite{Granovetter1973,Urena2020}. Sampling alters based on their log-weight ensures that higher-contact alters are also included in the null model. We estimate the expected number of triangles in a layer by taking 50 random samples of alters of size $n_L$ without replacement. Triangles from the aggregate network are sampled if their two conforming alters are randomly sampled. 
Figure~\ref{fig:triangles}c visualizes our triangle counts and Fig.~\ref{fig:triangles}d a comparison between observed and expected number of triangles. 
Triangles are more likely to be observed in constrained layers, i.e., when $n_L/k$ is small. 
This implies that the usage of social times is transitive when egos devote a certain social time to few of their alters: alters also use that time to communicate among themselves. Time usage is less transitive than in our null model when egos have broad social times, i.e., when their alter fractions $n_L/k$ are large. Such triangles still exist in the aggregate network, but the alters do not necessarily communicate in the same social time. A possible explanation in terms of focus theory is that when an ego devotes a social time to a small set of people, time usage captures small or similar foci; but when a social time is broad and not focalized to a small group, time usage captures either multiple or disjoint foci. We also find layer-specific behaviour: weekend layers tend to preserve triangles at a higher rate, whereas workweek layers tend under-represent triangles.

Last, we evaluate whether embedded alters in the personal network are more likely to exist within particular social times. Using the same sampling mechanism that assumes independence between alters and sampling weight proportional to log-contacts, we measure the number of triangles in the aggregate network constrained to the alters of a layer $T(A|L)$, a measure of the amount of embeddedness of the alters in the layer regardless of the social time. Figure \ref{fig:triangles}c depicts a visualization and Fig.~ \ref{fig:triangles}e results of this test. We find that well-embedded alters are more over-represented when the layer size is small $n_L/k$. In this case, such alters tend to be less over-represented during workweek layers. This suggests that these social times during the week are less indicative of focalized activity, or that communication on these layers likely includes collections of disjoint social foci. While triangles and well-connected alters do exist in these social times, they appear more frequently during the weekends. These results are consistent with those of Vergara Hidd \textit{et al.}, who found that stable relationships tend to communicate during the weekends \cite{VergaraHidd2023}. A possible explanation for this would be that stable relationships are also strong in the Granovetter sense and have overlapping circles or friends \cite{Granovetter1973}.

\subsection{Discussion}

Multiplexity manifests temporally through the use of different social times and has structural effects when representing such times as multilayer networks. We use weekly activity trends at a population level to infer latent social times, which act as a frame of reference for comparing the behaviour of individual ties. This serves two purposes, as it can both capture the individuality of ties in distributing their contacts, and the broad ways in which people use different social times. Ties display a very rich use of social times, including a broad spectrum of mono-multiplexity usage that can be untangled from contact volume and bursty behaviour.
We largely focused on temporal multiplexity to emphasize that it's the balancing of several layers that can affect patterns of bridginess or global connectivity, and not necessarily one particular layer. In this sense, people and connections may differ according to chronotypes and lifestyles, which can be reflected on how all social times behave as small and clustered social foci when egos use a social time for a small fraction of their alters. That being said, our results aligned with previous findings of the importance of weekends for strong and persistent ties, which tend to be embedded in circles of friends, and also that nighttime activity is associated with larger personal networks. 

Some of the limitations of our work are related to the generalizability of our results through our experimental design. Latent components are subject to stochastic fluctuations, a pre-specified number of components and orthogonality constraints. Our definition of temporal multiplexity that does not depend on particular social times can curtail some of these limitations, but they should be assessed separately when interpreting particular layers. In addition, we use a relatively old dataset from 2007, which is an asset from a sociological standpoint as it represents one of the major communication channels used at the time. However, the current trend of using diverse communication channels could affect the inference of latent components if people use different channels for different ties. More importantly, contemporary communication channels could impact social times themselves if they are, e.g., compound to demographic-specific social media use.

Our flexible framework couples social behaviour to network structure, offering a tool with several potential applications. From a theoretical perspective, this provides a new framework for modeling known communication patterns such as bursty behaviour. This has been previously analyzed under de-seasoning techniques \cite{Jo2012}, but to the best of our knowledge, not under the framework of tie-specific combinations of latent signals. A related question concerns the stability of usage of social times in large observation periods, which is likely related to the stability of relationships themselves \cite{VergaraHidd2023b}. Our method can be used to analyze other communication datasets and face-to-face interactions. Perhaps most importantly, it can shed light on other phenomena such the distribution of heterogeneous tie strengths in ego networks, aid in link prediction tasks, or provide more realistic models for epidemic and information spreading by representing social temporality in structural terms. 

\section{Methods}
\subsection{Non-negative matrix factorization}

We obtain latent population signals using a sample weekly activity matrix $X_s$ constructed from a random sample of $500,000$ rows from the full matrix $X$. 
We normalize the activity of each tie so that each bin contains the fraction of calls placed during that time instead of the total number of contacts, ensuring that high-contact ties are not over-represented. We obtain population-level latent components by decomposing the $500,000\times168$ non-negative weekly activity matrix $X_s$ via NNMF with orthogonality constraints \cite{Kimura2015}. 
The optimization problem is  $\min_{G,H} ||X_s-GH^T||^2_F + \lambda||H^TH - I||^2_F$, where $||*||_F$ is the Frobenius norm. $H\geq 0$ is a $168\times J$ matrix of social times, and $G\geq0$ is a $N_s\times J$ matrix of usage of social times by ties, or layer coefficients per tie. The second term represents the orthogonality constraint $H^TH\approx I_{J\times J}$ with its associated Lagrange multiplier $\lambda$. 
We obtain latent social times $H$ from a sample as NNMF can become computationally expensive for large matrices ($\sim 10$ million). 

Given the population signals $H$ we propose two methods to reconstruct ties in terms of the $J=4$ latent components. The first method $M1$ mimics the original NNMF problem, while method $M2$ uses reconstructs layers using a probabilistic representation. The latter allows to perform statistical model selection, overcoming the challenge of having competing signals that numerically approximate contact counts and tend to include values arbitrarily close to zero. Both methods use the tie-level vectors of total contacts per hour to retain a notion of call intensity. Most NNMF algorithms include stochastic components that find local optima and require multiple runs. We implemented the algorithm of Kimura \textit{et al.} \cite{Kimura2015} and performed 1000 runs with $J=4$ and selected the partition that minimized the orthogonality term. Setting the Lagrange multiplier $\lambda$ is an open problem in orthogonal NNMF \cite{Kimura2015}; we found that selecting models with maximal orthogonality resulted in signals that more closely resembled notions of social times across different $J$ values, see SM for the $J=4$ signal that minimized the approximation error, as well as different $J$ values of minimal orthogonality. 

\subsection{Data fitting}

\subsubsection*{M1: Layer weights from the original NNMF problem}

We restate the NNMF optimization problem by considering the component signal matrix $H$ as fixed. The new problem is $\min_G||X-GH^T||^2_F$, which can be simplified onto smaller tie-level optimization problems, allowing for parallelization and tractable numerical stability. 
Let $x$ and $g$ be the rows of $X$ and $G$ at the same index. We state the problem in the Euclidean norm $\min_g ||x^T - Hg^T||^2_2$. Eliminating terms that don't depend on $g$, the objective function follows the common quadratic programming statement: $\min_g gH^THg^T - 2xHg^T$ subject to $g\geq 0$. We used the quadratic programming solvers implemented on SciPy \cite{scipy2020}. 

\subsubsection*{M2: Layer weights using statistical inference}

We model the weekly activity profile of each tie as a multinomial random variable and use a maximum likelihood (ML) approach to find the parameter values. For simplicity, we omit the tie subindices $i$ used in the main text. We have that $X \sim \textrm{Mult}(w, h(\alpha))$, where $w$ is the number of trials (contacts, given by the data) and $h$ is a vector of probabilities parameterized as a linear combination of the columns of component matrix $\hat{H}$, a scaled $H$ where each column sums to 1. In other words, $h(\alpha) = \hat{H}\alpha$, where $\alpha \geq 0$ is a vector of coefficients that meets the condition $\sum_l \alpha_l = 1$. We restate the latter condition by defining $\gamma =\sum_l \alpha_l$, reparameterizing $\hat{\alpha} = \frac{1}{\gamma} \alpha$, and setting $\gamma \leq w$. This condition allowed for numerical stability when estimating a heavy-tailed array of $w$ values. 

For $x$ and $h(\alpha)$ vectors of $k=168$ observations and probabilities, respectively, the multinomial has the form $P_X(w, h) = \frac{w!}{x_1!\ldots x_k!} h_1^{x_1}\ldots h_k^{x_k}$, which results in a log-likelihood function $l(h) \propto \sum_i^k x_i \log(h_i)$. Expressed in terms of layer weights $\alpha$, the log-likelihood function includes a sum over the $J$ columns of $\hat{H}$, which represent the latent social times,  
\begin{equation}
    l\left(h(\hat{\alpha})\right)\propto \sum_i^{k=168} x_i \log\left(\frac{1}{\gamma} \sum_l^{J} \alpha_l \hat{H}_{il}\right).
\end{equation}
We find the ML estimates for $\alpha$ using numerical optimization. The gradient $\nabla l$ and Hessian matrix $\boldsymbol{H}_l$ can be constructed using
\begin{align}
    \frac{\partial l}{\partial\alpha_j} &= \sum_i^{k=168} \frac{x_i\hat{H}_{ij}}{\sum_l^{J} \alpha_l \hat{H}_{il}} - \frac{w}{\gamma}\\
    \frac{\partial l^2}{\partial\alpha_j\partial\alpha_m} &= -\sum_i^{k=168}\frac{x_i \hat{H}_{ij}\hat{H}_{im}}{\left(\sum_l^{J} \alpha_l\hat{H}_{il}\right)^2} + \frac{w}{\gamma^2}.
\end{align}

\subsection{Details on tests}

\subsubsection{Tie Multiplexity}

We test for temporal multiplexity using two reference models that account for the effects of contact volume and burstiness. First, we shuffle the weekly activity profiles according to the aggregate population behaviour and classify them using  the original latent signals $\hat{H}$. This results in shuffled activity profiles with no temporal correlations between bins, i.e., their layer coefficients can't be attributed to latent social timings. 
We shuffle each tie's $w$ contacts according to the aggregate population activity over the week. This means that if 1.5\% of all calls in the population were placed on Mondays between 14 and 15, or hour $j$, each of $w$ calls exists in that bin with probability $p_j=0.015$. In keeping with the multinomial notation, if $\tilde{X}$ is a $k=168$-sized random activity vector, then $P(\tilde{X}_j=1)=p_j$. 
We classify shuffled vectors of $w$ calls and obtain the distribution of induced multiplexity $S(\alpha|w)$ for a given weight. The focal multiplexity is the difference between the empirical, and the expected induced multiplexity $S(\alpha_i|x_i) - E[S(\alpha|\tilde{X}(w_i))]$. We estimate the latter term as the average value of shuffled vectors for a given $w$, where we include one shuffled tie per observed tie. 

We assess the sensitivity to burstiness in layer allocation. We compare the layer coefficients obtained using the data from the full activity profiles $\hat{\alpha}|x_i$ and the layer coefficients obtained from coarse-grained active hours $\hat{\alpha}|AH(x_i)$. The active hours $AH$ mapping takes value 1 when hour bin $j$ has at least one contact, and zero otherwise. We use JSD, a common measure of similarity between probability distributions, which takes value zero if the distributions are the same, and one if they differ fully \cite{Menendez1997}. 

\subsection{Bridginess}

We measured local bridigness using topological overlap over the aggregate network. Our dataset consists of users of mobile operator with a market share of 20\%. In line with previous work \cite{Urena2020}, we used an extended version of our dataset that includes calls to non-company users to compute overlap, as it contains complete ego networks around company users, computing overlap for ties where both nodes belong to the company. We tested global connectivity on the network of company users only. 

\subsubsection{Egos}

We tested for triangles around egos and the embeddedness of alters around egos. Both tests rely on counting the number of triangles around the alters observed in a layer at layer and aggregate levels. We used a random sample of 120K egos with minimum degree $k\geq 5$. We computed the layer-specific number of triangles around the ego $T(L) =\sum_{a\in L}t^L_a$. In other words, for each alter $a$ observed in layer $L$, and $t_a^L$ is the number of ties to other alters in the layer. We measured alter embededdness in a layer as the triangles count in the aggregate $A$ network, focusing on the alters selected in a layer $L$. In other words, $T(A|L)=\sum_{a\in L}t^A_a$, where $t^A_a$ is the number of ties to other alters in the aggregate network $A$. 

Our null tests reconstruct the social times of egos by taking random samples of the alters. Out tests assume that there are no correlations between alters, meaning that they use a social time to communicate with the ego independently of the other alters. However, we assume maximum transitivity, so that if two randomly-selected alters are connected in the aggregate, they are also connected in sample layer. This allows us to assess the relative use of social times, as it focuses on existing triangles in the aggregate network, and the degree to which those triangles are also present in the observed layers. For each ego with $n_L$ observed alters in layer $L$, we take 50 samples of alters of size $n_L$ without replacement from the aggregate networks. In other words, we obtain $\tilde{L}=\tilde{L}(n_L)$, a random sample of size $n_L$, and compute the expected number of triangles in a layer under random sampling $T(\tilde{L})=\sum_{a\in \tilde{L}}t^{\tilde{L}}_a$. We compare the ratio of observed and expected triangles $T(L)/E[T(\tilde{L}]$, estimating the expected triangles $E[T(\tilde{L})]$ using the average value of 50 samples $\tilde {L}(n_L)$. We account for the correlation between contact counts and topology by sampling each alter $a$ with probability proportional to the log-contact volume $\log(w_a)$. The embeddeness test uses the same principles, but does not need to account for transitivity. 

{\small \subsubsection*{Data and code availability}
The code used in this paper is available at 
\url{https://github.com/javurena7/social-foci/}. The CDR dataset analysed during the current study is not publicly available due to a signed non-disclosure agreement, as it contains sensitive information of the subscribers. 

\subsubsection*{Acknowledgments}
M.K. acknowledges support from the Academy of Finland grant numbers 349366, 353799, and 352561.
We acknowledge the computational resources provided by the Aalto Science--IT project. The study was part of the NetResilience consortium funded by the Strategic Research Council at the Academy of Finland (grant numbers 345188 and 345183).

\subsubsection*{Author contributions}
JUC, SH, TA, JS, and MK conceived, designed, and developed the study. JUC implemented and analyzed the models and empirical data studies, with SH performing additional analyses. JUC wrote the first draft and all authors revised and approved the manuscript. 

\subsubsection*{Competing interest statement}
All authors declare no competing interest.}

\bibliographystyle{science}

\end{document}